# History and Development of Coronal Mass Ejections as a Key Player in Solar Terrestrial Relationship

N. Gopalswamy[1]
NASA Goddard Space Flight Center, Greenbelt, Maryland, USA
[1]Corresponding author: nat.gopalswamy@nasa.gov

**Abstract**
Coronal mass ejections (CMEs) are relatively a recently-discovered phenomenon – in 1971, some fifteen years into the Space Era. It took another two decades to realize that CMEs are the most important players in solar terrestrial relationship as the root cause of severe weather in Earth's space environment. CMEs are now counted among the major natural hazards because they cause large solar energetic particle (SEP) events and major geomagnetic storms, both of which pose danger to humans and their technology in space and ground. Geomagnetic storms discovered in the 1700s, solar flares discovered in the 1800s, and SEP events discovered in the1900s are all now found to be closely related to CMEs via various physical processes occurring at various locations in and around CMEs, when they interact with the ambient medium. This article identifies a number of key developments that preceded the discovery of white-light CMEs suggesting that CMEs were waiting to be discovered. The last two decades witnessed an explosion of CME research following the launch of the Solar and Heliospheric Observatory mission in 1995, resulting in the establishment of a full picture of CMEs.

## 1. Introduction

In the history of solar terrestrial relationships, CMEs are a very recent phenomenon, discovered only in the beginning of the 1970s (Hansen et al. 1971; Tousey 1973). However, I show that CMEs were waiting to be discovered with some elements of CMEs were already known over the preceding decades. The purpose of this article is to provide an account of the discovery of various sub-structures of CMEs and the pre-eminent role played by space weather phenomena in finally obtaining the full picture of shock-driving CMEs, especially in visible light. Historical accounts on CMEs have been published a few times before (Kahler 1992; Gosling 1997; Howard 2006; Alexander et al. 2006; Webb and Howard 2012).

The wealth of knowledge accumulated on CMEs came from a series of coronagraphs: the first space borne coronagraph on board the seventh Orbiting Solar Observatory (OSO-7) mission (Tousey 1973), the Apollo Telescope Mount (ATM) Coronagraph on board Skylab (MacQueen et al. 1974), the Solwind coronagraph on board the Air Force satellite P78-1 (Michels et al. 1980), the Coronagraph/Polarimeter (CP) on board the Solar Maximum Mission (SMM) satellite (House et al. 1980), the Large Angle and Spectrometric Coronagraph (LASCO) on board the Solar and Heliospheric Observatory (SOHO) (Brueckner et al. 1995), the Solar Mass Ejection Imager (SMEI), a heliospheric imager on board the Coriolis spacecraft to observe interplanetary disturbances (Eyles et al. 2003), and the Sun Earth Connection Coronal and Heliospheric Investigation (SECCHI) on board the Solar Terrestrial Relations Observatory (STEREO) that includes the inner coronagraph COR1, the outer coronagraph COR2, and the two heliospheric imagers covering the entire space (Howard et al. 2008). Finally, the ground-based Mauna Loa K Coronameter (Fisher et al. 1981a) has been providing information on the corona close to the Sun. Publications based on observations from these coronagraphs will be used in highlighting key results on CMEs. We also use numerous other observations made from remote-sensing and in situ observations that uncovered a number of phenomena associated with CMEs and hence contributed significantly to the development of the full picture of CMEs known today. Excellent reviews on CMEs and the related phenomena already exist (Schwenn 2006; Chen 2011; Akasofu 2011; Webb and Howard 2012), so I focus only on the key milestones that led to the current understanding of CMEs.

Section 2 summarizes historical milestones that connected solar events to geomagnetic storms and cosmic ray modulation. The main space weather effect discussed –is the geomagnetic storm. Space weather implications of solar energetic particles became clear only in the space age.  Section 3 summarizes activities in the space era preceding the launch of SOHO.  Section 4 summarizes the exponential growth of knowledge on CMEs, aided by the continuous and uniform dataset obtained by SOHO. Results from the STEREO mission is also mentioned in this section. Finally, conclusions are presented in section 5.

**2. Early Days of Solar Terrestrial Relationship**
Herschel (1801) reported of five prolonged periods of few sunspots to be correlated with high wheat prices in England. He inferred that a fewer sunspots indicated less heat and light from the Sun so the wheat production was low resulting in high wheat prices. Herschel was ridiculed for this report, but now we know that when there are more sunspots, the Sun emits more radiation because of the brighter faculae appearing around sunspots. This discovery can be thought of as an early recognition of the periodic nature of sunspot activity, recognized half a century later by Schwabe in 1843. Coining of the term "geomagnetic storm" by Humboldt in the early 1800s synthesized the knowledge on geomagnetic disturbances observed since the mid1600s (see Howard 2006). Sabine (1852) found the synchronous variation of sunspot number and geomagnetic activity. In the backdrop of this long-term connection, Carrington and Hodgson independently discovered a sudden brightening on the Sun on September 1, 1859 and a huge geomagnetic storm was observed about 17 h later (Carrington 1859; Hodgson 1859). It took almost another half a century, back and forth, before connecting a geomagnetic event to a solar event (see discussion in Cliver 1994; Moldwin 2008), in particular, the ejection of particulate matter as a stream (FitzGerald 1892; Chapman 1918) or a plasma cloud (Lindemann 1919).

**2.1 Mass Emission from the Sun: Prominences and Flare Plasma**
We now know that prominences form the core of most CMEs. After the demonstration by Janssen and Lockyer in 1868 that prominences can be observed outside of eclipses using spectroscopes, the knowledge on prominences grew rapidly. By 1871, Secchi classified prominences into active and quiescent prominences. By 1892, it was a common knowledge that prominences erupted with speeds exceeding 100s of km/s (from Tandberg-Hanssen 1995). The disk counterpart of eruptive prominences were termed disparition brusque (DB) by d'Azambuja and d'Azambuja (1948).

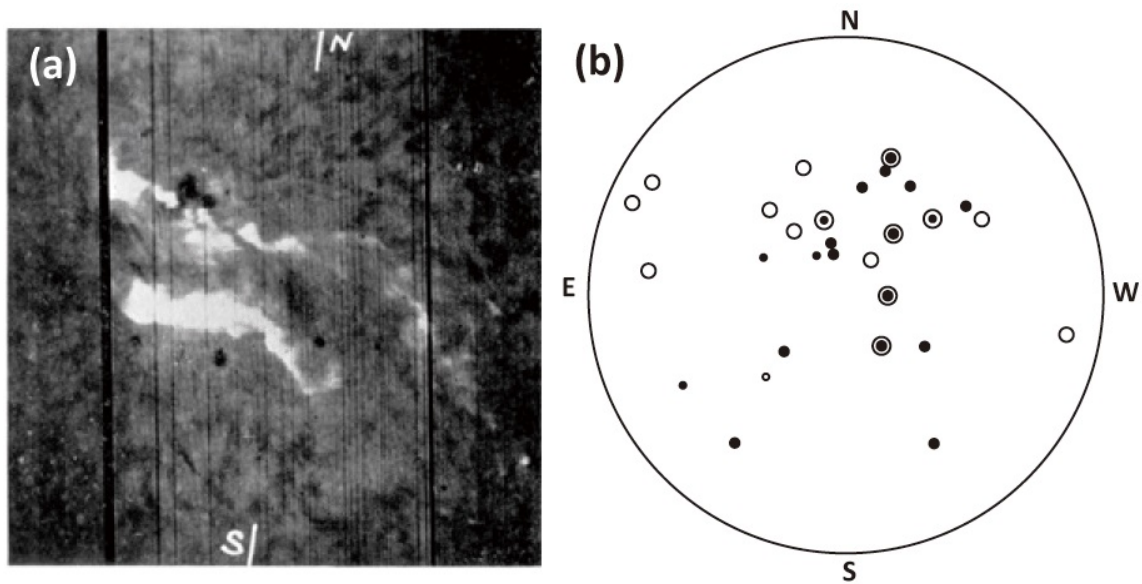

Figure 1. (a) The 1909 May 12 solar eruption photographed by G. E. Hale in H-alpha. A filament erupted from a sunspot group located at S13E15, which was a good location for an eruption to have direct impact on Earth (from Hale 1931). In modern terminology, this is a two-ribbon flare, typically formed after a filament eruption. Flares from locations like

this were followed by geomagnetic storms. (b) the open circles correspond to small storms, while the filled circles correspond to great storms. Encircled, filled symbols denote the outstanding storms that occurred over an 85-year period starting in 1859. (From Newton 1943).

Based on his compilation of great magnetic storms starting with the Carrington event, Hale (1931) found an average delay between the solar event and the geomagnetic storm to be about 26 hours. Figure 1a shows one of the eruptions reported in Hale (1931) which occurred on 1909 May 12 and was followed by a great storm ~33 h later. Shortly after this, Newton (1936) reported on observations that suggest the possibility of geomagnetic disturbances related to eruptive filaments (in H-alpha) outside of sunspot regions. Newton (1943) further developed this idea and established the flare – geomagnetic storm connection. In particular, he showed that the storm-causing eruptions need to occur close to the disk center – within ±45º in longitude and ±25º in latitude (see Fig. 1b). From these observations, he inferred that the angular extent of the corpuscular stream ejected at the time of the beginning of the flare was ~90º (similar to a modern CME extent) and that the transit time of the stream to Earth to be ~25.5 h. These numbers are very close to the modern CME values associated with intense geomagnetic storms.

**2.2 Energetic Particles from the Sun**
While the Sun–Earth relationship from the perspective of geomagnetic activity was under discussion for almost a century, energetic particles from the Sun were discovered only in 1942 (Forbush 1946). A large flare occurred near the center of the solar disk at the time of the solar cosmic ray event in 1946 on July 25, allowing Forbush to conclude on the flare origin of the particles. Before discovering the so-called solar cosmic rays, Forbush had already found the worldwide reduction in the intensity of cosmic rays (Forbush decrease) at the time of geomagnetic storms (Forbush 1937). In hindsight, it is easy to see that material ejections from the Sun is the cause of geomagnetic storms and Forbush deceases. Solar cosmic ray events observed on the ground are now known as ground level enhancement (GLE). Solar energetic particle events are significant source of space weather (Feynman and Gabriel 2000; Shea and Smart 2012).

**2.4 Radio Bursts and the Connection to Mass Emission and Shocks**
Payne-Scott et al. (1947) reported the observations of several solar radio bursts that showed a drift from higher to lower frequencies, which they interpreted as a physical agency moving away from the Sun with a speed of ~500-700 km/s similar to eruptive prominences. They also noted the association of the radio burst with a solar flare. These bursts were later classified as type II radio bursts (Wild and McCready, 1950).

Around this time, detailed theory of shock waves in infinitely conduction media had been worked out (de Hoffmann and Teller, 1950; Helfer 1953). Helfer (1953) even considered the possibility of shocks in solar prominences. Gold (1955) proposed that the storm sudden commencement (SSC) is caused by interplanetary shocks driven by gas emitted during solar flares. Type II bursts were also suggested to be due to MHD shocks (Uchida 1960). Morrison (1956) attributed cosmic ray modulation and Forbush decrease to large ionized gas clouds with tangled magnetic field ejected from the Sun. Parker (1957) worked out details on such magnetized gas clouds and showed that they can be ejected with speeds as high as the Alfven speed of the ambient medium. Parker used the term "magnetic cloud" to describe large scale structures with tangled magnetic fields, although this term is now used for flux rope structures with enhanced magnetic field, smooth rotation of the azimuthal component and low temperature (Burlaga et al. 1981).

Boischot (1957) discovered moving type IV radio bursts, which are magnetic structures carrying nonthermal electrons, and associated with solar eruptions. The radio source motion as fast as 1000 km/s or faster were reported (Boishot, 1958). This is can be regarded as the direct observation of magnetized material ejection from the Sun, apart from the eruptive prominences.

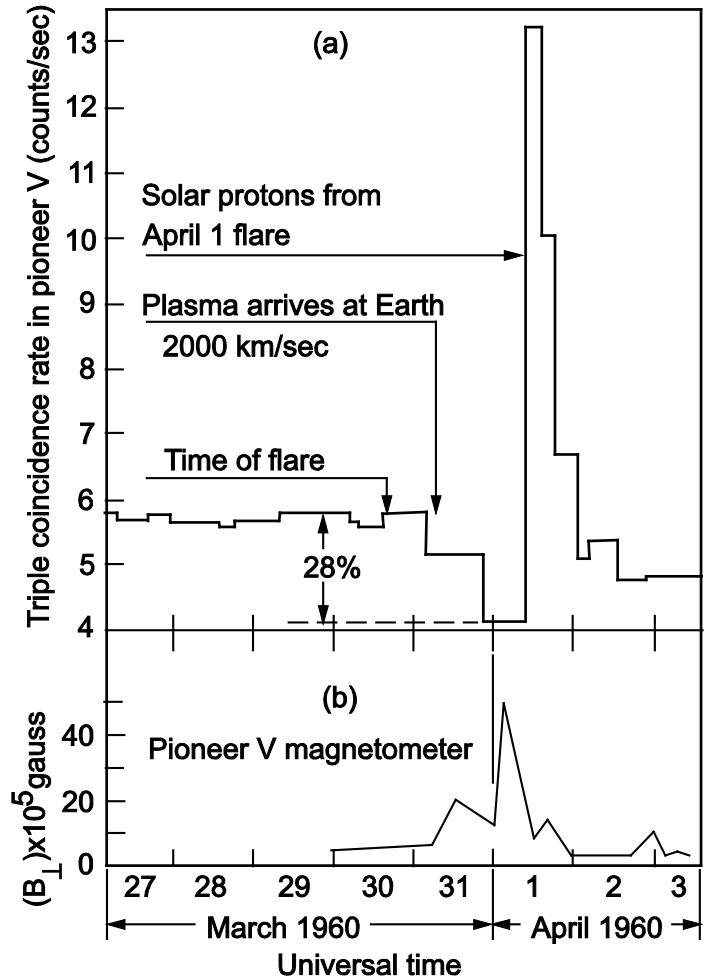

Figure 2. Particle (a) and magnetometer (b) data from Pioneer-V showing the reduction in cosmic ray intensity by 28% (Forbush decrease) associated with an interplanetary magnetic field enhancement. The increase in magnetic field in the beginning of March 31 is likely a shock, but the time resolution was not adequate to mark the shock (from Fan et al. 1960).

## 3. Advances Made during the Space Era
### 3.1 Interplanetary Origin of Forbush Decrease

The energetic particle and magnetic field measurements made by the Pioneer 5 mission during March-April 1960 (see Fig. 2) conclusively demonstrated that Forbush decrease is not due to the ring current but due to "conducting gas ejected at high velocity from solar flares" (Fan et al. 1960). This became a direct confirmation of Morrison's (1956) idea of cosmic ray modulation. The magnetometer data showed a magnetic field enhancement of about 50 nT coincident with the time of the Forbush decrease. A large energetic particle event (>75 MeV), probably due to another eruption was observed following the Forbush decrease. Fan et al. (1960) estimated that the speed of the flare plasma should have been

~2000 km/s. The Dst index associated with the arrival of the magnetic gas cloud was ~ -231 nT.

## 3.2 Direct Detection of an Interplanetary Shock

On 1962 October 7, Mariner 2 was located between the Sun and Earth, slightly behind Earth and above the ecliptic. A shock was detected at 15:42 UT and was followed by an SSC ~4.7 hours later (Sonnet et al. 1964). The SSC was not followed by a storm probably because just a flank of the shock impacted Earth. The shock detection and the associated SSC confirmed the suggestion by Gold (1955).

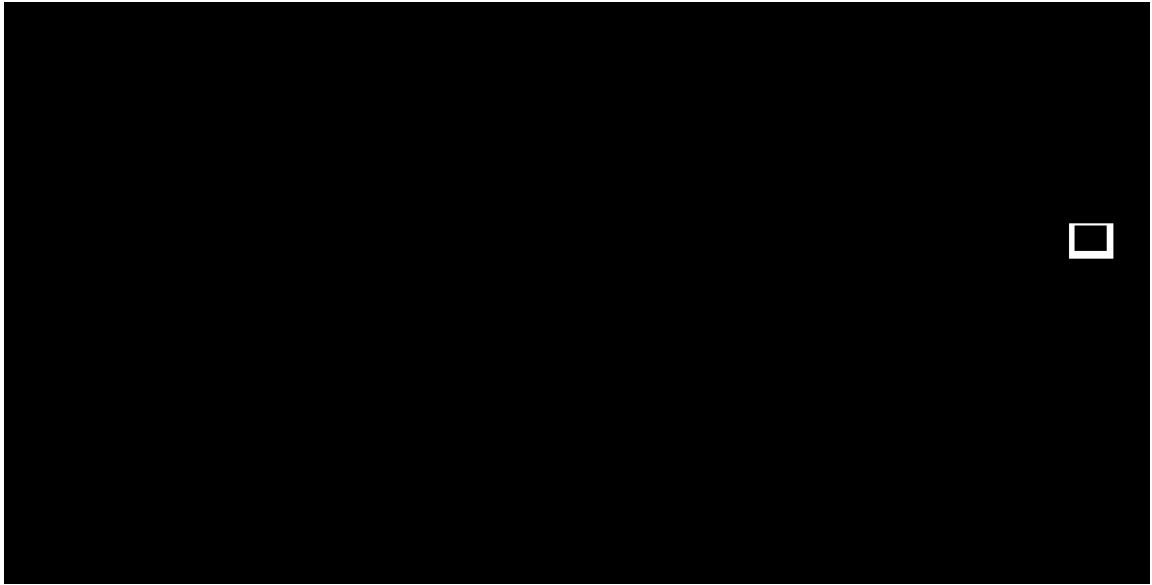

Figure 3. Magnetic gas cloud driving MHD shock as described by Fokker (1963) near the Sun (a) and by Gold (1962) in the interplanetary medium (b). The left picture (a) is based on the observation of type IV radio bursts caused by electrons trapped in stationary and moving magnetic structures. 'IVdecam' represents the type IV burst at decametric wavelengths. 'IVmB' denotes stationary type IV bursts due to electron trapped in loops that are quasi-stationary. The solar gas from eruptions was thought to have a frozen-in magnetic field that was described as a magnetic bottle. The MHD shock driven by the magnetic bottle is thought to be responsible for type II bursts. In (b), the shock standoff distance 'a' is determined by the radius of curvature of the magnetic bottle, the shock Mach number and the adiabatic index. The shock thickness 'd' is for collisionless shocks.

## 3.3 Early Picture of a Solar Eruption

By the early 1960s a complete picture of solar eruptions started emerging including the magnetized plasma ejections driving shocks, and their connection to geomagnetic storms. Gold (1962) developed an idealized structure of solar gas with embedded magnetic field connected to the Sun and driving shock at its leading edge (see Fig. 3). Fokker (1963) reviewed the properties of moving type IV bursts and their connection to geomagnetic storms and energetic particles from the Sun in the form of polar cap absorption of radio waves (due to 10-100 MeV particles precipitation in the polar region, Kundu and Haddock 1960) and GLE events (GeV particles reaching sea level). His schematic picture of the magnetic structure supporting the type IV burst and the leading shock producing the type II radio burst is very similar to the interplanetary picture developed by Gold (1962). It is worth noting that the picture in Fig. 3 is the same as the overall structure of fast CMEs we know today. With the establishment of the properties of the interplanetary medium consisting of the solar wind with Archimedean spiral magnetic structure (Parker 1958) and sector structure (Ahluwalia and Dessler 1962) it became clear that the magnetized plasma structures propagate through the background plasma. There was already considerable information on the density of the coronal plasma near the Sun based on radio observations (Ginzburg and Zhelezniakov 1958) and eclipse observations (Newkirk 1967).

**3.4 Energetic Particles, Shocks, and Plasma Clouds**
With the blast wave model of shocks from the Sun (Parker 1961), there was increased focus on particle acceleration by shocks (Weddell 1965; Hudson, 1965; Jokipii 1966). Bryant et al. (1962) reported the detection of a large increase in the intensity of 2-15 MeV protons by Pioneer 12 two days after the 1961 September 28 flare. The particle increase began just before the sudden commencement of a magnetic storm and hence termed 'energetic storm particle' event. Bryant et al. (1962) interpreted the event as protons from the Sun trapped within the plasma cloud that caused the magnetic storm. Examining similar events detected by the cosmic ray detectors on board Pioneer 6 and 7, Rao et al. (1967) concluded that the protons must have been accelerated by interplanetary shocks.

**3.5 A K-Coronal Transient from Ground-Based Observations**
The first white light observation of a coronal transient was observed during 1970 August 11-12 and reported by Hansen et al. (1971). A coronal streamer structure above the east limb suddenly disappeared and was accompanied by type II and type IV bursts (see Fig. 4). The 80 MHz type IV burst images obtained by the Culgoora radioheliograph coincided with the K-coronal transient. While the white-light transient was not tracked, the moving type IV burst was tracked and found to have a speed of 1300 km/s. Given the well-developed idea that type IV bursts are due to energetic electrons trapped in plasmoids with self-contained magnetic field (Smerd and Dulk, 1971; Schmahl, 1972), it became clear that the material ejection from the Sun manifests in a wide range of wavelengths.

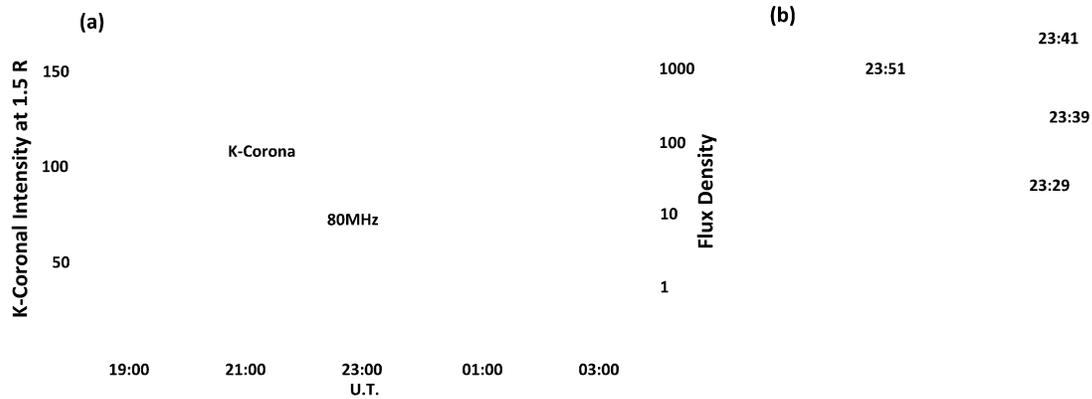

Figure 4. (a) Evolution of K-coronal intensity at a heliocentric distance of 1.5 $R_\odot$ above the east limb (dashed line) compared with the evolution of 80 MHz flux density (solid curve). (b) the position of the coronal transient feature is denoted by the solid curve (corresponding to 23:30 UT on 1970 August 11). The symbols indicate the moving type IV radio sources at different times. The open circle indicates the location of a stationary radio source. The solid and dashed circles are at heliocentric distances of 1 and 1.5 $R_\odot$, respectively (from Hansen et al. 1971).

Malitson et al. (1973) were also able to track an IP shock -from the Sun to 1 AU using a type II radio burst observed by the IMP-6 mission. Watanabe et al. (1973) used the interplanetary scintillation (IPS) technique to identify transient interplanetary disturbances associated with solar flares. The transients were identified as an enhancement in the solar wind speed over a very short duration (see also Rickett et al. 1975). Gosling et al. (1973) found that the ejecta behind interplanetary shocks detected by Vela 3 had unusually low proton temperature, high speed, low density and helium enrichment. These authors suggested that the low temperature must have resulted from the magnetic bottle configuration in the ejecta and the adiabatic cooling of the plasma in the bottle. Gosling et al.'s (1973) observations provided direct confirmation of the Gold (1962) picture shown in Fig. 3.

**3.6 The First Coronal transient from Space: OSO-7**
On 1971 December 14 the white-light coronagraph on board NASA's OSO-7 mission detected and tracked a transient (Tousey 1973). During the mission life of OSO-7 (September 29, 1971 – July 9, 1974), twenty three coronal transients were observed. Koomen et al. (1974) identified the white-light transient with the Gold bottle, while Stewart et al. (1974) confirmed the picture in Fig. 3(a) complete with the leading shock ahead of the white-light cloud and a moving type IV burst immediately behind the cloud followed by prominence material in the deep interior of the cloud. Figure 5 shows the height-time history of various features of the coronal transient from radio (type II, type IV bursts), white light (K coronal transient), and H-alpha (prominence material).

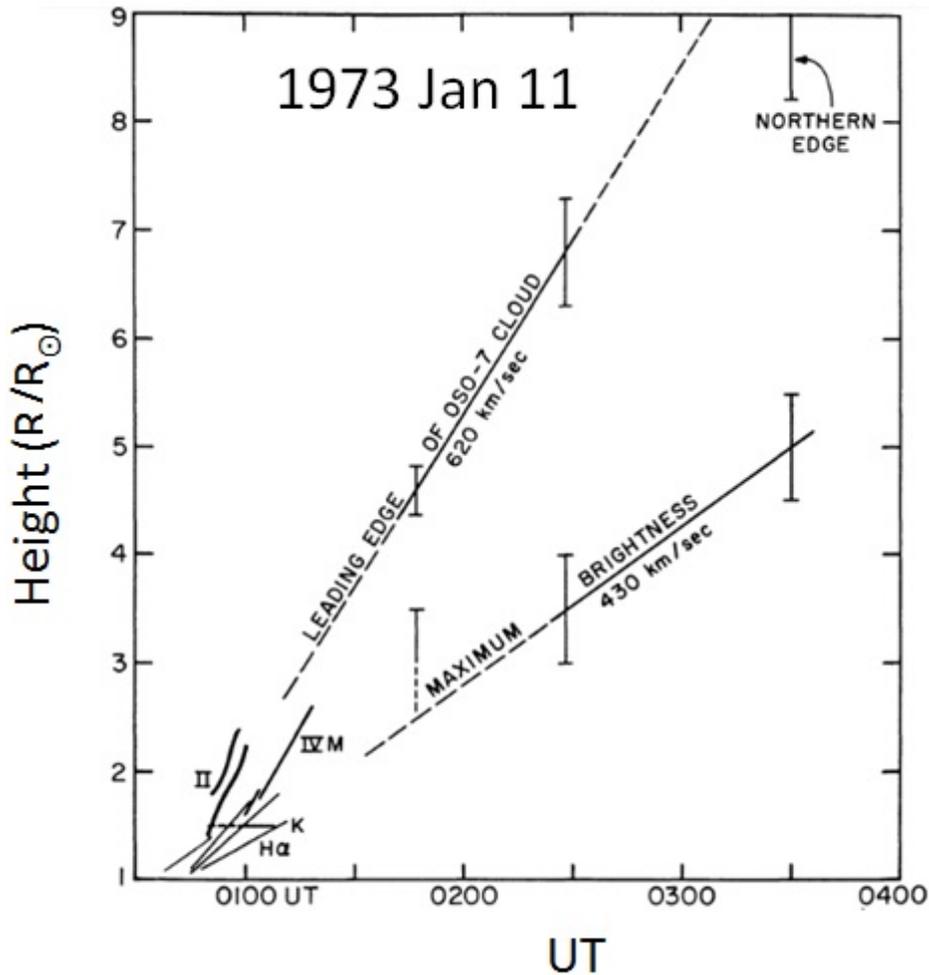

Figure 5. The height-time history of various substructure of a white light coronal transient observed on 1973 January 11 by OSO-7. The type II and type IV bursts imaged by the Culgoora Radioheliograph are ahead and behind the K-coronal transient, while the bright H-alpha material is at the bottom of the cloud (from Stewart et al. 1974)

Soon after the report on the detection of the OSO-7 coronal transient, Eddy (1974) identified a coronal transient in the eclipse observations of the 1860 July 18 eclipse. Cliver (1989) pointed out that what was thought of as a comet during the 1893 April 16 eclipse is actually a CME event. Webb and Cliver (1995) searched a hundred years of eclipse observations looking for CMEs. There have also been other CME detections during modern eclipses (Rušin et al. 2000; Hanaoka et al. 2014).

**3.7 Skylab CMEs**
The Skylab Apollo Telescope Mount (ATM) Coronagraph that had operations overlapping with OSO-7 (May 1973–Feb 1974) detected 110 coronal transients in white light over a period of 227 days and led to a number of key results. The transients also became known as coronal mass ejections around this time. Gosling et al. (1975) was the first to identify a Skylab CME with an interplanetary (IP) shock: the 1973 September 7 CME from the southwest quadrant of the Sun (S18W46) resulted in the shock observed by Pioneer 9 located at 0.98 AU directly above the source region on the Sun on September 9. The shock and the driving ejecta at Pioneer 9 had speeds of 722 and 600 km/s, respectively. The CME speed near the Sun was ≥960 km/s, indicating a clear deceleration in the IP medium. The CME mass was also estimated to be ~$2.4 \times 10^{16}$ g. Another key result was the significant correlation between the occurrence rate of CMEs and the sunspot number (Hildner et al. 1976). Kahler et al. (1978) studied 16 Skylab CMEs and found a good correlation between the CME speed and SEP intensity. These authors concluded that "energetic protons are accelerated in the shock front just ahead of the expanding loop structures observed as mass ejections." This result formed the basis for the establishment of shock acceleration as the leading mechanism for SEPs (Cliver et al. 1982; Cane et al. 1988; Gosling 1993; Reames 1995). Mouschovias and Poland (1978) suggested that some CMEs can be thought of as a "twisted rope of magnetic-field lines expanding and broadening in the background coronal plasma and magnetic field" and modeled the 1973 August 10 Skylab transient as a rope. Based on the polarization measurements of the same CME, Crifo et al. (1983) suggested that the transient must be a three-dimensional bubble-like structure. Fisher and Munro (1984) introduced the ice-cream cone model of CMEs to explain a Mauna Loa Mark III CME observed on consistent with Crifo et al. (1983).

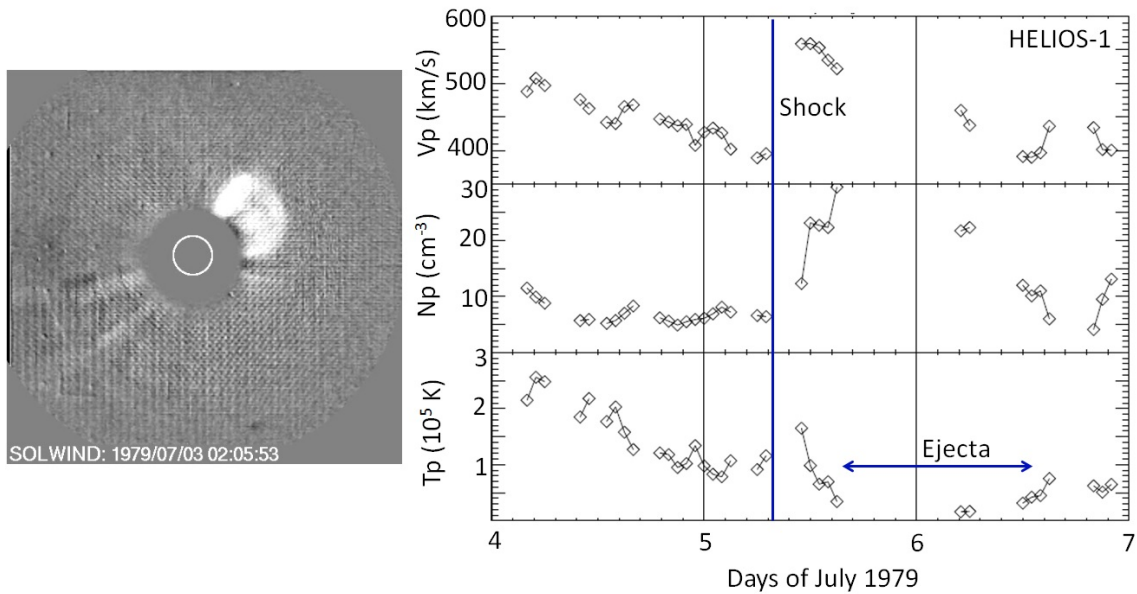

Figure 6. The 1979 July 3 CME observed by Solwind (left) and the IP shock driven by an ejecta observed on July 5 by the HELIOS-1 spacecraft (right). Tp, Np, and Vp are the plasma temperature, density and the flow speed of the solar wind. The approximate time of the IP shock and the extent of the ejecta are marked. Note that the ejecta marks the region of low proton temperature (Gosling et al. 1973). Helios-1 was above the west limb of the Sun, so the part of the CME remote-sensed by Solwind was intercepted by HELIOS-1, thus making it possible to study the slowdown of the CME between the Sun and 0.84 AU.

**3.8 The Solwind CMEs**
When the Solwind coronagraph on board the P78-1 mission became operational (February 1979 to September 1985), it observed ~1700 CMEs. The Solwind observations confirmed many of the Skylab results and led to a number of new discoveries. Halo CMEs were identified as CMEs directed toward Earth appearing to surround the occulting disk in sky-plane projection (Howard et al. 1982). Solwind data also indicated rich morphological features in CMEs (Howard et al. 1985). Sheeley et al. (1980) reported a high-latitude CME and correctly inferred that such CMEs must be associated with polar crown filaments observed in large numbers during solar maxima. High-altitude CMEs represent the process by which the old magnetic flux is removed from polar regions, paving the way for the appearance of the new magnetic cycle according to the hypothesis by Low (1997). By combining Solwind CME observations with in-situ observations from HELIOS and Pioneer Venus Orbiter (PVO), it was possible to identify the interplanetary acceleration of CMEs (Lindsay et al. 1999) and quantify it for CME travel time predictions (Gopalswamy et al. 2001). Sheeley et al. (1985) found that almost all IP shocks were associated with CMEs near the Sun, confirming the case study of Gosling et al. (1975). Figure 6 shows a Solwind CME that was intercepted by HELIOS-1 after 61 hours as a shock-ejecta system. The ejecta is the region with the lowest proton temperature (Gosling et al. 1973). Near the Sun, the CME had a speed of ~582 km/s. HELIOS-1 was located above the west limb of the Sun at a distance of 0.84 AU and hence was in quadrature with P78-1. The speed of the ejecta measured in-situ was ~470 km/s, indicating a clear deceleration of the CME.

The recognition that a halo CME is a regular CME propagating in the Earthward or anti-Earthward direction (Howard et al. 1982) is an important step in understanding how CMEs impact Earth. Howard et al. (1985) studied about 1000 Solwind CMEs (1979-1981) and found that there was no obvious correlation between CME rate and the sunspot number contradicting Hildner et al. (1976) results from Skylab. However, when the full set of data was used in combination with data from other missions, the CME rate was found to track the solar activity cycle in both amplitude and phase (Webb 1991; Webb and Howard 1994). Kahler et al.'s (1978) Skylab result that energetic particles are due to CME-driven shocks was also confirmed with a larger sample (Kahler et al. 1983). Cliver et al. (1983) associated the 1979 August 21 GLE with a Solwind CME with a sky-plane speed of ~700 km/s (real speed is likely to be >1000 km/s, when projection effects are taken into account).

### 3.9 The Solar Maximum Mission CMEs

The Coronagraph/Polarimeter (CP) on board SMM observed CMEs in 1980 and 1984-1989, adding another 1200 CMEs to the CME data base.  SMM was interrupted in 1980 when the satellite's attitude control system failed. In 1984, the crew of space shuttle Challenger (STS-41-C) captured, repaired, and redeployed the SMM mission, which continued operations until 1989.

The SMM/CP observations abundantly confirmed the three-part structure of CMEs (frontal structure, dark void, and prominence core - Fisher et al. 1981b) and the solar sources were found to be closely associated with sites of filament eruptions (Hundhausen 1993). Because of the quadrant field of view with a limited radial extent (1.5 to 6 Rs), SMM/CP observed generally slower CMEs. SMM/CP also did not observe many halo CMEs because it took about 90 min for obtaining a whole Sun image. Howard (1987) compared 18 CMEs observed by both SMM/CP and Solwind and found that the CMEs showed a tendency to accelerate within the combined field of view (1.5 to 10 Rs). Thus Solwind would record a larger speed for the same CME due to the acceleration. The SMM data also did not show a strong solar cycle variation (see Fig. 7). Although there are peaks in the mean speed around the solar maximum epochs (1981-1982, 1989), the strong peak during the minimum phase in 1985 makes it difficult to see the solar cycle variation (see also Ivanov and Obridko 2001).  The mean CME speed of SMM CMEs in 1985 (458 km/s) was significantly higher than that of Solwind CMEs (150 km/s). It turned out that these speeds correspond to different sets of CMEs observed in 1985 by the two coronagraphs and that the speed of nearly half of the SMM CMEs were not measurable (Gopalswamy et al. 2003a).

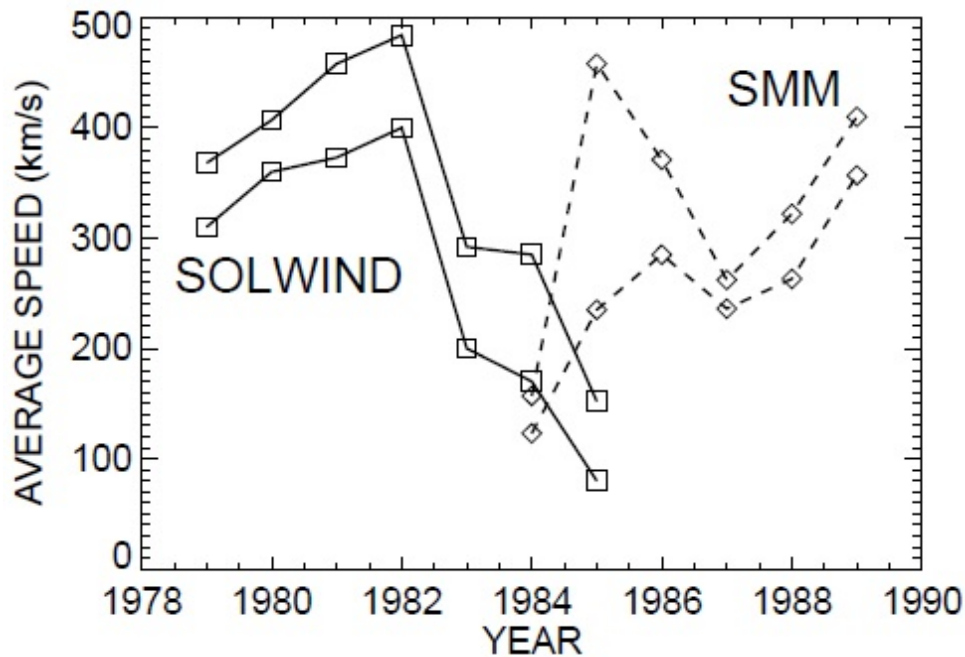

Figure 7. Annual averages of CME speeds from Solwind (solid curves) and SMM/CP (dashed curves) data. The upper and lower cures correspond to mean and median values of yearly data (from Gopalswamy et al. 2003a).

During the SMM era, the Clark Lake multifrequency radioheliograph imaged the thermal free-free (thermal bremsstrahlung) emission from the CME plasma at metric wavelengths, confirming the enhanced density in the frontal structure (Gopalswamy and Kundu 1992) and the mass estimates from coronagraph observations. Detailed comparison between CMEs observed by the ground based Mauna Loa K-corona meter and SMM/CP revealed that most CMEs accelerated within the SMM/CP FOV (St.Cyr et al. 1999). Kahler (1994) identified fast SMM/CP CMEs (>1300 km/s) associated with three GLEs. Analysis of these events revealed that the GeV particles were released when the CMEs were in the height range 3-5 Rs. By studying a set of truly limb CMEs, Burkepile et al. (2004) showed that projection effects significantly affect the statistical properties such as kinematics, mass, and occurrence latitude of CMEs.

### 3.10 Other Heliospheric Observations
In addition to natural radio sources, scintillation of satellite signals have also been used to study shocks. Woo and Armstrong (1981) used the radio signals from Voyager 1, 2 to detect a fast (~3500 km/s) shock at a distance of 13.1 Rs from the east limb of the Sun. Using the IPS technique, Gapper et al. (1982) identified solar disturbances to heliocentric distances beyond Earth's orbit. Propagation of solar disturbances into the interplanetary medium was also inferred from HELIOS zodiacal photometer data as a short-term increase in polarized light. Richter et al. (1982) reported on eight clouds and estimated the size along the line of sight to be ~0.06 AU (see also Jackson 1985; Webb and Jackson 1990). Neutron monitors (Bieber and Evenson 1997) and muon detectors (Munakata et al. 2005) in the worldwide network remain an important source of information on Forbush decrease and GLEs. Low frequency radio receivers also provide important information on CME-driven shocks in the inner heliosphere because the emission frequency indicates a heliocentric distance of the shock (Cane et al. 1987; Bougeret 1997).

### 3.11 Shock Sheath and Flux Rope
An important refinement of the Gold picture in Fig. 3b was made by Burlaga et al. (1981) who identified the region between the shock and the magnetic cloud as the turbulent hot plasma compressed by the shock and consisting of highly variable magnetic field. Burlaga et al. (1981) used plasma and magnetic field data from five spacecraft (Helios-1,2, IMP-8, Voyager-1,2) to derive the 3-D structure of the driver consisting of high magnetic field, low temperature and density, and rotation of the field vectors indicating a loop. They called this structure a magnetic cloud consisting of helical magnetic field, subsequently known as flux ropes (Burlaga and Klein 1980; Klein and Burlaga 1981).

The helical fields inferred from interplanetary observation is consistent with the magnetic ropes at the Sun suggested by Muschovias and Poland (1978), although the connection was not made at the time. These observations are consistent with the magnetic bottle inferred earlier (Fig.3), except that the magnetic field is twisted. Goldstein (1983) modeled the magnetic structure of magnetic clouds as a flux rope based on the twisted cylindrical fields considered by Lundquist (1951) in conducting fluids.

### 3.12 Magnetic Clouds and Southward Magnetic Field Component
The critical connection between CMEs and geomagnetic storms was made by Wilson (1987), who found that the southward component of the interplanetary magnetic field in magnetic clouds is responsible for the storms. Wilson (1987) performed a superposed epoch analysis of 19 magnetic clouds and found that the Dst index simultaneously decreased to a large negative value at the onset of a large, sustained southward magnetic field in the magnetic cloud; the storm recovery started when the magnetic field turned northward. The results were consistent with the well-known connection between geomagnetic disturbances and the southward component of the external (interplanetary) magnetic field (Dungey 1961; Fairfield and Cahill 1966). Zhang and Burlaga (1988) extended this study to various types of clouds. Southward interplanetary field can also occur in the shock sheaths (Gonzalez and Tsurutani 1987; Gosling and McComas 1987), which have been shown to be equally important in causing geomagnetic storms (Tsurutani et al. 1988).

### 3.13 CMEs as Primary Space Weather Driver
By late 1980s it became clear that the southward component of the magnetic field in the interplanetary CMEs (ICMEs) and/or the shock sheath is directly responsible for intense geomagnetic storms (Wilson 1987; Gonzalez et al. 1987). It also became clear that CME-driven shocks are responsible for the large SEP events, although flares can accelerate particles that can be distinguished by the higher Fe/O ration (Reames 1990). However, flares continued to be considered as the primary source of shocks and CMEs (Maxwell and Dryer, 1982). The term "flare-generated shock" continued to be used giving secondary importance to the CME (see e.g., Dryer 1975; Woo and Armstrong 1981). On the basis of a broad spectrum of evidences, Kahler (1992) concluded that CMEs play a primary role in accelerating SEPs and producing IP shocks. Gosling (1993), pointed out that the emphasis on flares is misplaced as far as geomagnetic storms and large SEP events are concerned and he referred to this as solar flare myth. Gosling's (1993) conclusions represented the culmination of research on CMEs in the 1970s and 1980s bringing them to the forefront as the main players in affecting Earth's space environment.

## 4. CMEs in the SOHO Era

SOHO was launched in 1995 and has been observing the corona up to a distance of 32 Rs over the past twenty years. SOHO's uniform and continuous data set has provided clarity to many of the results obtained from the past coronagraphs, and led to many new discoveries. An order of magnitude more CMEs have been observed than the combined set of all CMEs observed in the pre-SOHO era (http://cdaw.gsfc.nasa.gov/CME_list, Yashiro et al. 2004; Gopalswamy et al. 2009a). Thousands of papers have been published using SOHO data, so it is impossible to list all the results. We highlight some key results that are relevant to space weather.

### 4.1 CME Rate and Correlation with Sunspot Number

The fact that the CME rate is correlated with sunspot number (SSN) reported from pre-SOHO coronagraphs (Webb and Howard 1994) was readily confirmed with SOHO CMEs for cycle 23 (Gopalswamy et al. 2003a). When the correlation was considered separately for the rise, maximum, and declining phases, it was found that the maximum phase had the weakest correlation (Gopalswamy et al. 2010a). This was attributed to the fact that the number of CMEs associated with quiescent filament eruptions (i.e., non-spot) from the high and low latitudes increases during the maximum phase (Gopalswamy et al. 2003b,c). This might be the reason that Howard et al. (1985) did not find good correlation between the CME rate and sunspot number for the period 1979-1981, which is the maximum phase of cycle 21.

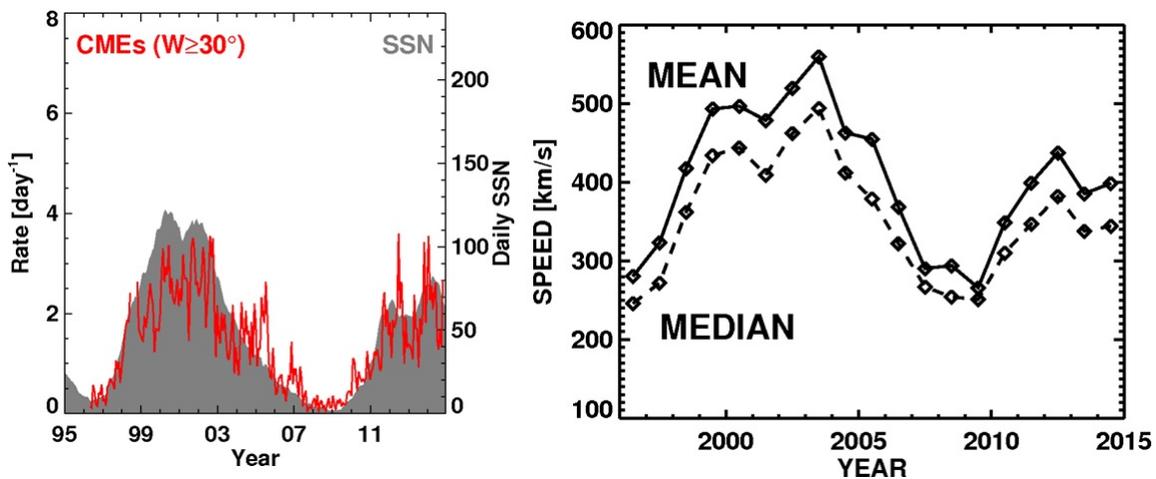

Figure 8. (left) SOHO CME rate superposed on sunspot number (SSN, gray). The CME rate is averaged over Carrington Rotation periods. (right) Yearly-averaged mean and median speeds of SOHO CMEs.

Figure 8 shows the close correlation between the sunspot number and the CME rate. We can also see major deviations during the maximum phase. Intra-cycle variation can also be seen in the CME rate: the CME rate relative to the Sunspot number is clearly larger in cycle 24 (2009-2015). The average speeds also show a clear solar cycle variation, including the double peak in the maximum phase (the double peak indicates peaking of the sunspot activity at different times in the two hemispheres). This is a clear confirmation of the speed variation, which was not clear in the pre-SOHO era (see Fig. 7). The weaker solar activity cycle 24 seems to have important consequences for CMEs in the heliosphere: CMEs expand anomalously due to reduced heliospheric pressure leading to the increased observed rate of small CMEs, halos originating far from the disk center, and mild space weather (Gopalswamy et al. 2014; 2015a; Petrie 2015).

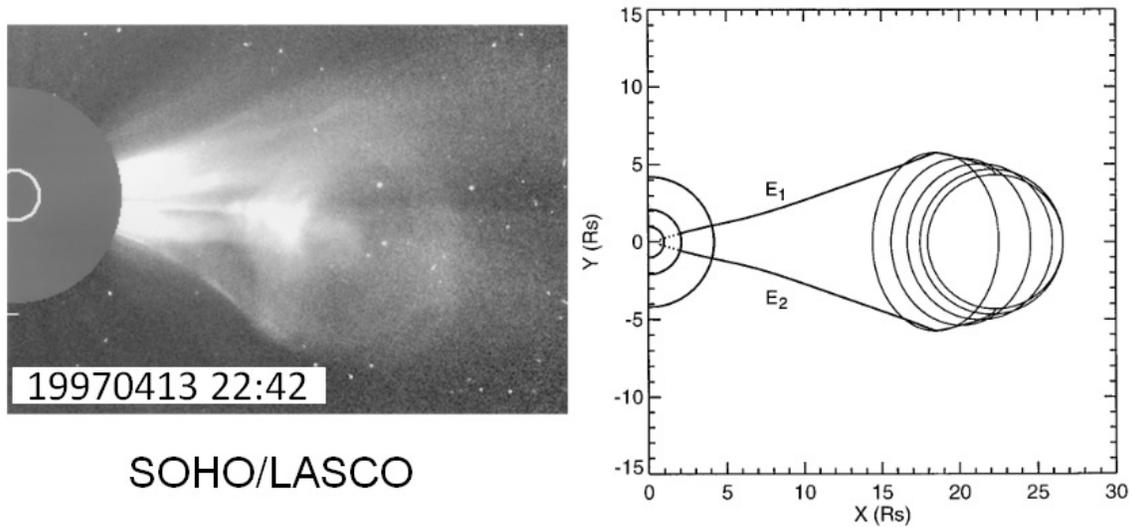

Figure 9. A flux rope CME from SOHO LASCO and the model (from Chen et al. 1997).

**4.2 Flux Rope CMEs**
A magnetic flux rope can be thought of as a locally cylindrical bundle of field lines that have a helical structure except for the axial field line. The field strength has a maximum value on the axis and decreases towards the edges of the cylinder. The rope geometry proposed by Muschovias and Poland (1978) can be thought of as the first suggestion that some CMEs can be modeled as magnetic flux ropes. SOHO provided direct evidence of flux ropes in coronagraph images. Figure 9 shows a flux rope CME observed in the LASCO field of view on 1997 April 13, which was modeled by Chen et al. (1997). These authors confirmed the flux rope morphology based on the agreement between the observed CME expansion and the theoretically calculated flux-rope dynamics. Most CME modeling work involves flux ropes, which may be present before eruption or formed during eruption (see Gibson et al. 2006; Linton and Moldwin 2009 for reviews). There is considerable evidence to support the idea that all ICMEs may be flux ropes (Marubashi 1997; Owens et al. 2005; Gopalswamy 2006; Riley et al. 2006; Gopalswamy et al. 2013a; Vourlidas et al. 2013; Marubashi et al. 2015). Therefore, flux ropes ejected at the Sun do appear as flux ropes in the IP medium, suggesting that the flux ropes are not destroyed during their propagation. Observationally, some CMEs appear as non-ropes because the measuring spacecraft may be passing through the legs of the flux ropes or flanks of the ICME, thus not detecting the flux rope structure. One of the important implications for space weather is that identification of a flux rope near the Sun makes it easier to predict the expected magnetic configuration at Earth (Burlaga and Klein 1980; Burlaga et al. 1981; Goldstein 1983; Lepping et al. 1990; Qiu et al. 2007) modified by the interaction with the solar wind on the its way to Earth.

**4.3 Halo CMEs**

Halo CMEs are so called because the excess brightness in these events appear to surround the occulting disk of the observing coronagraph as projected on the sky plane. Halo CMEs were discovered in Solwind images (Howard et al. 1982) but remained a novelty because only a handful were observed over the entire lifetime of Solwind. The SMM/CP did not observe many halos because most halos are observed at distances beyond ~6 Rs, which may be the reason why many halos were observed by SMM/CP and also SMM observed in quadrants which meant a lower cadence for the whole corona. SOHO has observed several hundred halo CMEs in the past 20 years, indicating that about 4 % of all CMEs are halos (Gopalswamy et al. 2010b). Frontside halos (those originating on the Earth-facing side of the Sun) are important for space weather because they are expected to directly impact Earth as magnetic clouds (Webb et al. 2000; St. Cyr 2005; Gopalswamy et al. 2007, 2010b). Halo CMEs are similar to normal CMEs, except that they are faster and wider and hence more energetic. For example, there were 696 fast ($\geq$900 km/s) and wide ($\geq$60°) CMEs observed during January 1996 to March 2015. Over the same period, 662 halos were observed, of which 319 were fast. Some halos with speed <900 km/s may be faster in reality because of projection effects. With SOHO images alone, one does not know the intrinsic width of halo CMEs. When SOHO was in quadrature with one of the two STEREO spacecraft, it was possible to show that the STEREO limb CMEs were halo CMEs in SOHO and that they were wide (typical width >60°) and the speeds were in the range (510–2200 km/s) (Gopalswamy et al. 2013b). The fraction of halo CMEs in a CME population has been found to be a good indicator of the average CME energy in the population (Gopalswamy et al. 2010a).

**4.4 CME Propagation: Deflection, Interaction, and Rotation**

A systematic equatorward deflection of CMEs observed during 1973–1974 by ~2° was reported in the pre-SOHO era (Hildner 1977; MacQueen et al. 1986). SOHO images abundantly confirmed this, but helped quantify the extent of CME deflection by the enhanced magnetic field in coronal holes. Gopalswamy and Thompson (2000) found that both the prominence and the CME showed the deflection, suggesting that the CME deflected as a whole, including substructures. Furthermore, the deflection was over ~30°, much larger than what was reported in the pre-SOHO era. Deflections were also observed in the longitudinal direction due to the magnetic field of equatorial coronal holes. In fact the deflection could be in any direction, depending on the relative position of the coronal hole and the eruption region (Cremades et al. 2006; Gopalswamy et al. 2001). For space weather effects, deflections relative to the Sun-Earth line are important (Gopalswamy et al. 2009b). Deflections toward the Sun-Earth line can make a CME more geoeffective, while a deflection in the opposite direction can make a CME miss Earth. Coronal-hole deflection can also result in the lack of alignment between the CME and shock (Wood et al. 2012) if the shock is weakly driven. The physical reason for the deflection has been shown to be due to the global magnetic pattern surrounding the eruption region (Filippov et al. 2001; Sterling et al. 2011; Gui et al. 2011; Panasenco et al. 2013; Kay et al. 2013; Möstl et al. 2015).

CME-CME interaction in the SOHO coronagraph field of view was reported by Gopalswamy et al. (2001). The interaction caused a change in trajectory of the preceding CME and a broadband enhancement of the associated type II burst occurred suggesting additional electron acceleration. Higher SEP intensity results when there is CME-CME interaction (Gopalswamy et al. 2004) and the arrival of CMEs at 1 AU gets delayed when they interact with preceding CMEs (Manoharan et al. 2004; Temmer et al. 2012). CME interaction can also result in a single shock at Earth even though multiple CMEs are ejected at the Sun (Gopalswamy et al. 2012a).

CME Rotation is somewhat controversial. It has been known for a long time that the orientation of interplanetary flux ropes are generally aligned with the polarity inversion line (or filament) at the Sun (Marubashi 1997; Bothmer and Schwenn 1998; Yurchyshyn et al. 2001; Marubashi et al. 2009). However, several authors have interpreted CME observations to indicate rotation during coronal and interplanetary propagation. Yurchyshyn et al. (2009) fitted ellipses to the outlines of halo and partial halo CMEs and compared them with the axis of the associated post-eruption arcades. They reported that CMEs appear to rotate by about 10° for most of the events with about 30–50° for some events (see also Isavnin et al. 2014). Vourlidas et al. (2011) reported an event with a rotation rate of 60° per day. Recently, Marubashi et al. (2015) analyzed a set of more than 50 well observed CME-ICME pairs and found strong support to the idea that an erupted flux rope has its main axis parallel to the polarity inversion line and remains so as it propagates through the interplanetary space. This is also consistent with models in which the flux rope is formed due to reconnection (Kusano et al. 2012). A clear definition of rotation is needed in three dimensions, for example, with respect to the radial direction from the source region.

**4.5 Flare-CME Relationship**
Although CMEs are responsible for space weather effects, they can be still closely related to flares: the eruption process has been understood to be common to both the mass ejection and flare heating (see e.g. Gosling 1990). SOHO observations support the idea that the flare and CME are two manifestations of the same energy release (e.g., Low 1994; Harrison 1995), although the CME kinetic energy has been found to be the single largest component of the released energy (Emslie et al. 2012). The connection between flares and CMEs has been recognized in a number of ways: coupled evolution indicated by the similarity of CME speed profiles to that of flare soft X-ray emission (Zhang et al. 2001), significant correlation between CME kinetic energy and soft X-ray fluence (Burkepile et al. 2004; Gopalswamy 2009), flare origin of enhanced charge state in ICMEs (Reinard 2008; Gopalswamy et al. 2013a), CME nose located radially above the flare site (Yashiro et al. 2008; Veronig et al. 2010), and the correlation between reconnection flux and the azimuthal flux of IP flux ropes (Longcope and Beveridge 2007). It is well known that CMEs do occur with almost no surface signatures (Wagner 1984; Robbrecht et al. 2009), suggesting that all the released energy must be going into mass motion. However, flare energy measured as the thermal energy content of even extremely weak post-eruption arcades from the polar CMEs, seem to have quantitative relationship with CME kinetic energy (Gopalswamy et al. 2015b). The flare-CME relationship is somewhat muddled when CMEs erupt from complex magnetic regions involving active regions and filaments (Harrison et al. 2012).

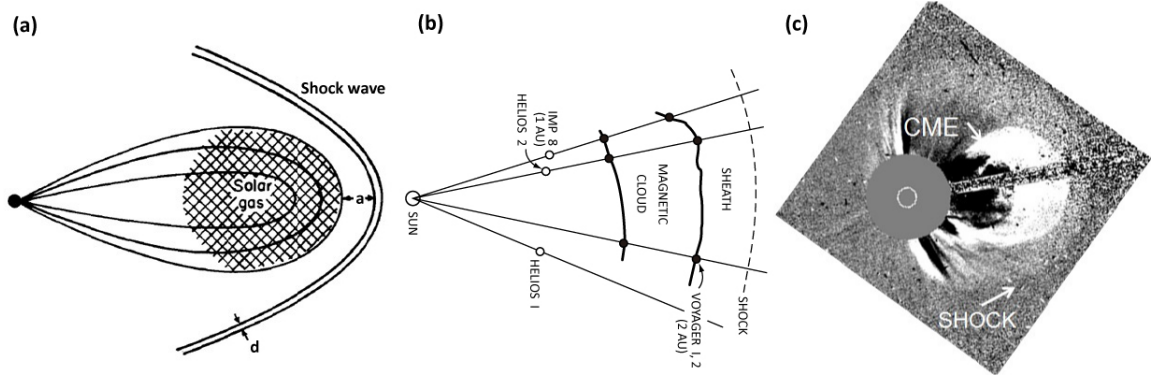

Figure 10. (a) An interplanetary shock driven by a solar magnetic structure as envisioned by Gold (1962) in order to explain sudden commencement of geomagnetic storms. (b) An IP shock driven by a magnetic cloud (MC) with a sheath between the shock and MC as inferred from multi-spacecraft observation by Burlaga et al. (1981). (c) A white-light CME from SOHO/LASCO with a diffused sheath region surrounding the CME from Gopalswamy (2010). Note that the magnetic structure with entrained solar gas (a), magnetic cloud (b), and CME (c) all refer to the flux rope that drives the shock.

**4.6 Detection of shocks using white-light imaging**
Although a shock ahead of CMEs is inferred from spectral and imaging observations of type II bursts (see Fig. 3), it was not observed in white light images. Gosling et al. (1976) suggested that MHD bow waves ahead of CMEs must form shocks if the CMEs are fast enough. Shocks were inferred based their impact on nearby coronal structures. Only in the SOHO era was the shock structure discerned: Sheeley et al. (2000) identified the shock feature surrounding the bright CME material as "the disturbances are faintly visible ahead of the ejected material at the noses of the CMEs but are strongly visible along the flanks and rear ends…. these disturbances are shock waves…". The correspondence of Gold's (1962) shock with the magnetic cloud-sheath-shock system inferred from multi-spacecraft observation (Burlaga et al. 1981), and a white-light shock surrounding a CME observed by SOHO (Gopalswamy 2010) is shown in Fig. 10. It must be pointed out that the shock itself is too narrow to be resolved by the coronagraph so, the diffuse feature ahead of the flux rope is the shock sheath and the outer edge of this sheath is taken as the shock location. The brightness of the sheath region is caused by the density jump across the fast mode MHD shock. This completes the full picture of a shock driving CME conceptualized by Gold (1962), identified in the IP medium by Burlaga et al. (1981) and finally identified in SOHO white light images by Sheeley et al. (2000).

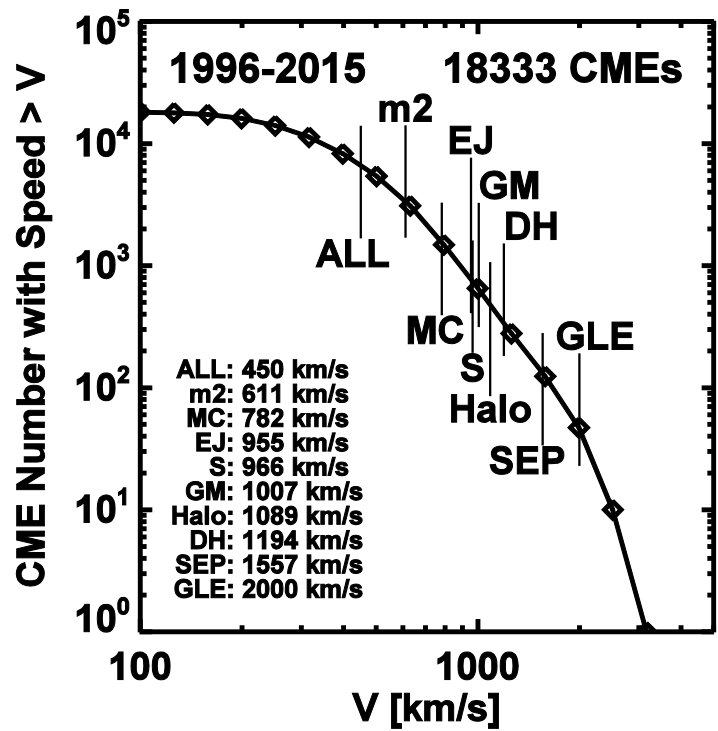

Figure 11. Cumulative distribution of CME speeds (V) for the period 1996-2015. The average speeds of various subsets of CMEs are marked. ALL - the general population, m2 – CMEs associated with metric type II bursts without interplanetary extension, MC – CMEs associated with magnetic clouds detected at 1 AU, EJ – CMEs associated with non-cloud ICMEs detected at 1 AU, S – CMEs associated with interplanetary shocks at 1 AU, GM – CMEs associated with major geomagnetic storms (Dst $\leq$-100 nT), Halo – halo CME population, DH – CMEs associated with decameter-hectometric (DH) type II radio bursts, SEP – CMEs associated with large SEP events, and GLE – CMEs associated with ground level enhancement (GLE) in SEP events.

## 4.7 CME Relevance for Space Weather

Since SOHO has observed more than 20,000 CMEs in the past 20 years, it is worth looking at what fraction of these CMEs are important for space weather. Figure 11 shows the cumulative distribution of CME speeds for the period 1996-2015. The speeds of all CMEs are not measurable, so the number of CMEs in Fig. 11 is less than the total number. The average speeds of various subsets of CMEs is marked on the cumulative speed plot. The average speed of the general population is ~450 km/s, which is essentially the same as what was obtained using all pre-SOHO data (Gopalswamy 2004). The maximum speed exceeds 3000 km/s, but only in a couple of cases. About 50 CMEs had a speed exceeding 2000 km/s (Solwind detected only one CME with a speed of ~2000 km/s). Gopalswamy et al. (2010b) suggested that one does not observe CMEs with speeds exceeding ~4000 km/s because of the limit on the maximum free energy that can be stored in solar active regions. The limit was estimated to be ~$10^{36}$ erg assuming a hypothetical active region with the largest area ever observed containing the largest magnetic field strength ever observed in sunspots. A single CME usually does not exhaust all the available free energy. The largest kinetic energy observed was $1.2 \times 10^{33}$ erg for the Halloween 2003 CME on October 28 (Gopalswamy et al. 2005a).

All the subsets of CMEs have average speeds above that of the general population. CMEs producing purely metric type II bursts are of lowest speed (611 km/s), which is sufficient to drive shocks near the Sun (within about 1 Rs from the surface). CMEs accelerating SEPs with ground level enhancement are the fastest (~2000 km/s). SOHO CMEs were identified with all GLE events of cycle 23 and found to be the fastest subset driving shocks throughout the inner heliosphere (Kahler et al. 2003; Gopalswamy et al. 2005b; Cliver 2006). CMEs associated with decameter-hectometric (DH) type II busts and SEP events are similar because the same shock is responsible for both the phenomena. CMEs associated with magnetic clouds (MCs), ejecta (EJ), IP shocks (S), and geomagnetic storms (GM) represent plasma impact on Earth while the others represent particle acceleration by CME-driven shocks. All subsets except for those associated with MCs have speeds close to 1000 km/s or higher. MCs are associated with CMEs from close to the disk center, so their true speed may be higher. Thus we see that only about 1000-2000 CMEs (or <10%) are important for space weather.

## 4.8 STEREO Contributions

The two additional views provided by the twin STEREO spacecraft helped understand the three-dimensional structure of CMEs (Howard et al. 2008). For example, the same CME observed in two views confirmed that halo CMEs are regular CMEs but viewed head-on (Gopalswamy et al. 2010a). STEREO instruments helped us observe CMEs much closer to the Sun (see e.g., Gopalswamy et al. 2009c; Patsourakos et al. 2010) and all the way to 1 AU (Harrison and Davies 2014 and references therein) because of the appropriate instrument suite includes the inner coronagraph COR1 and the heliospheric imagers (HIs). CMEs were tracked in three dimensions (Mierla et al. 2010) all the way to 1-AU using the HI images (see e.g., Möstl et al. 2010; DeForest et al. 2013; Liu et al. 2014). DeForest et al. (2013) exploited the availability of quantitative photometry from STEREO/SECCHI to distinguish between mass picked up from the wind and mass that is part of the original CME. HI observations (from SMEI and STEREO) have helped improve the prediction of arrival times of CMEs (Webb et al. 2009) and provide advance warning of storm-causing CMEs. In particular, the arrival of the CME-driven shock can be viewed in HI images, normally inferred from solar wind plots (Möstl et al. 2010). It was also possible to obtain the height of GLE release directly from STEREO/COR1 images (Gopalswamy et al. 2013c; Thakur et al. 2014), confirming earlier results from Kahler (1994). Observations close to the Sun also helped quantify the initial acceleration of CMEs (Bein et al. 2011), which was also possible using the Mauna Loa K Coronameter images (Fisher 1984; St Cyr et al. 1999; Gopalswamy et al. 2012b). Multiple views also helped recognize many SEP events that are observed all around the Sun (see e.g., Dresing et al. 2012; Lario et al. 2013) providing opportunities to test the CME origin and heliospheric propagation of energetic particles.

## 5. Summary and Conclusions

Coronal mass ejections are a novelty because they were discovered only in 1971, some 15 years into the Space Era. It took another 30 years and half a dozen space missions to obtain the complete picture of CMEs. The discovery of CMEs can be traced to the attempts by scientists to understand the cause of geomagnetic disturbances, Forbush decreases, and solar energetic particles. Simply because flares were discovered first, it was natural to identify flares as the cause of the interplanetary disturbances. Yet, people realized early on that it was matter rather than electromagnetic radiation that caused large geomagnetic and interplanetary disturbances. A shock was proposed to explain the storm sudden commencement and energetic storm particles. A shock was also needed to explain type II radio bursts. The driver was identified to be a magnetized plasma structure, although initially thought to be just a plasma cloud. The last thing to be identified in the corona is the white-light shock.

Space weather effects can be directly attributed to CME sub-structures. While the outermost structure (shock) accelerates particles and causes sudden commencement, the interior structures (sheath and flux rope) cause geomagnetic storms if they possess a strong southward component of the magnetic field. It has become clear that CMEs are a natural hazard because they produce space weather effects that pose danger to human technology in space and ground: from telegraph systems to GPS navigation; from power grids on the ground to payload in Mars orbit. We still have a long way to go in predicting when a CME will occur on the Sun, and when it will arrive at a given destination in the heliosphere and what the magnetic structure would be. We still need to figure out the relative contribution of flares and CMEs to the observed energetic particles from solar eruptions.

**Acknowledgments:** I thank R. Selvakumaran for checking the references and S. Akiyama for improving some figures. I also thank V. Bothmer, J. Chen, C. E. DeForest, R. A. Harrison, Y. Liu, B. C. Low, K. Marubashi, M. Mierla, C. Möstl, V. N. Obridko, D. V. Reames, A. Reinard, O. C. St. Cyr, A. Sterling, N. Thakur, and D. F. Webb for helpful comments. This work was supported by NASA's LWS and TR&T programs. This work benefitted greatly from the open data policy of NASA.

**Competing interests**
The author declares that he has no competing interests.